\documentclass[10pt]{revtex4}
\usepackage{graphicx}
\usepackage{amsmath,amssymb}
\usepackage{mathtools}

\begin{document}

\title{Universal quantum theory from dynamical consistency}
\author{Chiara Marletto and Vlatko Vedral}
\affiliation{Clarendon Laboratory, University of Oxford, Parks Road, Oxford OX1 3PU, United Kingdom}

\begin{abstract}
We argue for the universality of quantum theory using a dynamical consistency argument, within a specific Hamiltonian setting. We analyse two different types of coupling between simple quantum harmonic oscillators. Each illustrates an aspect of the free and interacting quantum fields and shows the inadequacy of semiclassical models. In particular, we establish that requiring the canonical algebra to be preserved under joint unitary dynamics rules out specific hybrid classical-quantum models.
 We apply our reasoning to the gravitational field in the linear regime, coupled to the quantised electromagnetic field and, separately, to quantised matter. We conclude with a comparison to DeWitt's analysis of quantum measurement, in which the apparatus, if classical, must be at least stochastic to preserve the Heisenberg Uncertainty Principle. We also note that stochastic models are inconsistent with the strict version of conservation principles, even if they comply with a probabilistic (on average) conservation.
\end{abstract}

\maketitle

One of the most striking aspects of quantum theory is that its information-theoretic structure cannot be fully captured by classical models unless one gives up basic properties such as locality or locally held conservation laws. Phenomena such as contextuality, Bell nonlocality, and Heisenberg's uncertainty principle -- or the impossibility of copying all states of a quantum system -- are key signatures of chiefly quantum systems. This fact has motivated extensive efforts to characterise the boundary between classical and quantum models, both from foundational and operational perspectives. Hereinafter by ''classical" we shall mean a commutative algebra of observables represented by c-number variables (possibly stochastic), rather than a Hilbert-space operator algebra.

A common approach to studying this boundary is through quantum-classical hybrid models, in which some degrees of freedom are treated quantum mechanically while others remain classical. Such models arise naturally in diverse contexts, ranging from semiclassical gravity and quantum control to measurement theory and open-system dynamics. However, consistently coupling quantum and classical systems has proven to be remarkably difficult. Attempts to combine quantum and classical dynamics often encounter fundamental inconsistencies, including violations of positivity, failure of energy or probability conservation, or incompatibility with the statistical structure of either theory, \cite{SALC1, SALC2}. More recently, no-go theorems have shown that under broad and physically motivated assumptions, hybrid quantum--classical models cannot reproduce the full range of quantum phenomena without sacrificing essential physical principles -- such as locality or conservation laws, \cite{MV-njp, PERTE, OPPE, MAVE}. Understanding the origin of these inconsistencies provides valuable insight into both the nature of quantum non-classicality and the constraints that any viable extension or approximation of quantum theory must satisfy.

The purpose of this work is to explore interacting systems and how the consistency of the resulting formalism forces us to quantise the whole system if one of the subsystems is already quantum. Such arguments have been employed since the beginning of quantum physics, but here we present them in a modern setting and emphasise the wide range of constraints that the resulting dynamics must obey. We shall consider hybrid descriptions in which one sector is represented by operators on a Hilbert space, the joint dynamics is generated by a Hamiltonian, and the canonical commutation relations are required to be preserved exactly. Under these assumptions, nontrivial interactions of the forms considered below are incompatible with a purely classical representation of the second sector.

We start by revisiting Heisenberg's argument for the quantisation of the electromagnetic field \cite{Heisenberg}. The basic idea is that since charges are sources of the electromagnetic field, it is inconsistent to treat charges quantum mechanically while leaving the field in classical form. In fact, if a classical field could faithfully couple to quantum charges, one could violate the Heisenberg uncertainty principle by ``copying" the position and the momentum of the charge into the electric and magnetic field components. 

Indeed, the equation for an oscillator of frequency $\omega_0$ and position $x$ driven by the time-dependent electric field $E(t)=E_0 e^{i\omega t}$ is given by:
\[
\ddot x + \omega_0^2 x = \frac{e}{m}E_0 e^{i\omega t} \; ,
\]
which can be solved to obtain
\[
x(t) = \frac{e}{m(\omega_0^2-\omega^2)}E_0 e^{i\omega t}\propto ae^{-i\omega t} + a^\dagger  e^{+i\omega t}\; ,
\]
where we have now expressed the field in terms of the creation and annihilation operators.
From here, the momentum follows by differentiation
\[
p(t) = \propto -i (ae^{-i\omega t} - a^\dagger e^{+i\omega t})
\]
leading us to conclude that 
\[
[x(t),p(t)] \propto [E(t),B(t)] \; .
\]
This makes it clear that it is inconsistent for the material oscillator (representing a charge) to be quantised while the field that drives it remains classical. Similar arguments have been used by Milonni to show that the quantum vacuum mode density must take a specific form for the radiation-reaction formula to remain valid \cite{Milonni}. We note various models of DeWitt \cite{DeW-Book}, in which DeWitt considered a general quantum measurement and examined the backreaction of the system on the apparatus arising from the measurement interaction. We have also previously considered hybrid systems \cite{MV-njp,MV-njpq} and their capacity for processing different kinds of information. 

We now investigate two forms of coupling between harmonic oscillators, which are among the most fundamental systems in physics. The reason for this is that all quantum fields are ultimately collections of simple quantum harmonic oscillators. We do so with the intention of extending the above logic to other relevant interactions.  

The following Hamiltonian can either describe modes of free fields or of two fields coupled to each other:
\[
H = \hbar \omega (a_1^\dagger a_1 + a_2^\dagger a_2) + \hbar g (a_1^\dagger a_2 + a_2^\dagger a_1)
\]
Heisenberg's equations of motion are given by  \cite{Narducci}
\[
\frac{d}{dt}
\begin{pmatrix} 
a_1  \\ 
a_2 
\end{pmatrix}
= -i
\begin{pmatrix} 
\omega & g \\ 
g & \omega 
\end{pmatrix}
\begin{pmatrix} 
a_1  \\ 
a_2 
\end{pmatrix}
-i 
\begin{pmatrix} 
0 & g \\ 
g & 0 
\end{pmatrix}
\begin{pmatrix} 
a_1^\dagger  \\ 
a_2^\dagger
\end{pmatrix}
\]
Solving Heisenberg's equations of motion, one obtains:
\[
x_1 (t) = \cos (\omega t) \cos (gt) x_1(0) - \sin (\omega t) \sin (gt) x_2(0)+\frac{1}{\omega}\sin (\omega t) \cos (gt) p_1(0)+\frac{1}{\omega}\cos (\omega t) \sin (gt) p_2(0)
\] 
and similarly for $x_2(t)$. From these, we can obtain the equations of motion for $p_1 = m dx_1/dt$ and $p_2=mdx_2/dt$. This allows us to compute the equal-time commutators between positions and momenta of the two oscillators.
\[
[x_1(t),p_1(t)] = \cos^2 (gt) [x_1(0),p_1(0)] + \sin^2 (gt) [x_2(0),p_2(0)]
\]
This suffices to make the following point. If oscillator $1$ is treated quantum mechanically, then so must be oscillator $2$. Otherwise, the above equation is inconsistent because the commutation relations for oscillator $1$ cannot be conserved in time. 
Let us illustrate this with an example. The coupling Hamiltonian between the electromagnetic and gravitational field is, in the linear approximation, given by:
\[
H_I \propto h_{ij}T_{ij}
\]
where we assume that the stress tensor is due to the magnetic component of the field
\[
T_{ij} \propto (B_i+b_i)(B_j+b_j)
\]
where $B$ is the constant (classical component of the field) while $b$ designates the dynamical part, and 
\[
h_{ij} \propto (a_{ij}+a^\dagger_{ij})
\]
The resulting Hamiltonian now has the form considered above:
\[ 
H_I \propto B_ib_j(a_{ij}+a^\dagger_{ij})
\]
It is therefore clear from the previous argument that if the magnetic field is quantised (i.e., if $b_j = \beta e^{-i\omega t} + \beta^\dagger  e^{+i\omega t}$), the gravitational field must be quantised as well. This provides a dynamical-consistency argument against treating the gravitational field as a purely classical field in this setting, even though the effects of quantising gravity may well be very small in practice \cite{Dyson}.

Now we look at a different Hamiltonian, also relevant for interacting fields
\[
H = \hbar \omega (a_1^\dagger a_1 + a_2^\dagger a_2) + \hbar g a_1^\dagger a_1(a_2+ a_2^\dagger)
\]
The Heisenberg equations of motion for the creation and annihilation operators are:
\begin{eqnarray}
\frac{da_1}{dt} & = & -i ( \omega + g (a_2+ a_2^\dagger))a_1\\
\frac{da_2}{dt} & = & -i ( \omega a_2 + g a_1^\dagger a_1)
\end{eqnarray} 
whose solutions are given by 
\begin{eqnarray}
a_1 (t) & = & e^{-i\phi(t)}D(\alpha (t))a_1(0)\\
a_2 (t) & = & e^{-i \omega t} a_2(0) + \alpha^* a_1^\dagger a_1
\end{eqnarray} 
where $\phi (t) = \omega - g^2/\omega^2 t$, $D(\alpha (t))$ is the displacement operator acting on the second oscillator and $\alpha = g/\omega (e^{i\omega}-1)$. 

For our purposes, it will suffice to look at the unequal time commutator between the annihilation operators for the two oscillators:
\[
[a_1(t),a_2(t')] = \alpha^* a_1(t)
\]
(Similar results are obtained for all other commutators). It is clear that if $a_1$ is represented by a q-numbers, so must be $a_2$ for otherwise the LHS would be equal to zero, while the RHS would be an operator.  

We now illustrate the same idea using the case of linear gravity coupled to quantised matter, and using the assumption of locality (no action at a distance). The relevant interaction Hamiltonian is:
\[
H_I \propto b^\dagger b (a+a^\dagger)
\]
which could describe a mass coupling to the linear gravitational field \cite{MAVE,SOUG}. Again, the above analysis suggests that if the mass is quantised, so must the field, if one is to avoid violation of the basic unequal time commutation relations for the massive particle. 

What about the static interaction in the field? Could one not just write a direct, classical interaction between the masses of the kind $(b^\dagger b)^2$? One argument against this possibility is that the resulting Hamiltonian is non-local -- because the two masses are directly coupled via the classical field. However, the resulting dynamics obeys microcausality. Therefore, the direct non-local interactions still comply with relativity. However, if you insist that the non-local interaction must be obtained from the local interaction, then this forces the quantisation of the field. For that to be possible, one needs a unitary transformation that maps the local interaction to a non-local one, thereby eliminating the field. Such a unitary transformation does not commute with $H_I$: this ultimately requires that the field be quantised. We can see this via this argument: 

\begin{eqnarray}
e^{iS}He^{-iS} & = & (1 + iS-\frac{S^2}{2}+...)H(1 - iS-\frac{S^2}{2}+...) \\
& = & H + i[S,H] + \frac{1}{2}[S,[H,S]] +...
\end{eqnarray}
In order to cancel the interaction, we need that 
\[
[S,H] = i\hbar g b^\dagger b (a+a^\dagger)
\]
so that 
\[
S= i \frac{g}{\omega}b^\dagger b (a+a^\dagger)
\]
The second commutator then gives us the direct interaction:
\[
\frac{1}{2}[S,[H,S]] = -\frac{1}{2} (b^\dagger b)^2
\]
The above calculation shows that the commutator $[a,a^\dagger]$ must have the unit value, as otherwise the local Hamiltonian could not be converted into the non-local one. 
Although this kind of transformation $S$ is normally performed as part of renormalisation to remove infrared divergences (by removing the field and redressing the charges), here we see it as a generator of a gauge transformation that takes us from the Lorenz gauge to the Coulomb gauge. Whatever the motivation, we have concluded that the field needs to be treated quantum mechanically for that transformation to be possible. A charge simply cannot be dressed by a classical field because the charge could be in a superposition of different states, which automatically forces the field to be able to do the same if dressing is to be achieved. This concludes this further argument which holds under the locality assumption.

As a final note, we would also like to point out that there is another argument, based on conservation laws, to refute the semi-classical field model. An exact conservation law for an additive quantity $O=O_A+O_B$, expressed on a bipartite system consisting of two sectors $A$ and $B$, takes the following form. The only allowed unitaries $U$ on the sectors $A$ and $B$ are those with the property that 
\begin{equation}
[ O, U]=0 \;, \label{cons}
\end{equation}
This exact conservation law leads, as pointed out in \cite{MAVEcons}, to a conservation in a `branch-by-branch' sense. If one expands the quantum state of the composite system in the basis of eigenstates of $O_A$ and $O_B$, even when $O_A$ or $O_B$ are locally unsharp, the allowed interactions preserve the quantity $O$, so that if in a branch $O_A$ increases on $A$, $O_B$ must decrease in value on $B$ to compensate, and vice versa. When one of the sectors (say $B$) is a classical system, the exact conservation law becomes problematic. There are various ways to represent the classicalisation of $B$. One is to assume $B$ has only observables that commute with $O_B$ (including its allowed local states $\rho_B$). This would mean that the allowed unitaries on $B$ are trivial and cannot change $\rho_B$: this scenario does not correspond to a physically viable description. Another possibility is that $B$ is described by a stochastic model, expressed as a c-number stochastic variable $o_B$. One could then require that an average conservation law hold, with the conserved quantity $\langle O_A\rangle + o_B$, where $\langle O_A\rangle$ is the quantum expected value of $O_A$. This would be a much weaker requirement than the operator identity \eqref{cons} and would not allow for the branch-by-branch conservation law. Therefore, committing to a classical model for a field coupled to quantum matter is tenable only if one also gives up on exact conservation laws. This move appears problematic, as most known symmetries are expressed as exact conservation laws rather than as averages. 

In this work, we have examined the issue of classical-quantum hybrid systems adopting a dynamical consistency argument. We have found once more that consistent hybrid models are impossible given a set of plausible physical assumptions. Our analysis further supports the view that genuinely quantum behaviour cannot, in general, emerge from or interact consistently with a purely classical substrate without introducing non-classical features into the latter. We note that we have assumed unitarity, a standard assumption in other arguments of this kind \cite{SALC1, DeW-Book}. Our conclusions concern hybrid models satisfying three assumptions: (i) the quantum sector obeys the standard operator algebra, (ii) the combined dynamics is generated by a Hamiltonian and is unitary, and (iii) the interaction is local. We do not claim to exclude all conceivable nonunitary, nonlinear, stochastic, or operationally defined quantum-classical theories; rather, we show that such theories cannot retain all of these assumptions simultaneously. It would be interesting to make this work more general by dropping this assumption, using, for instance, other information-theoretic frameworks as in \cite{MV-PRD}.
Beyond their foundational significance, these results have important implications for ongoing efforts to describe systems at the interface between classical and quantum physics, including semiclassical gravity, quantum control, and effective theories of measurement. Rather than motivating increasingly sophisticated hybrid models, the recurring inconsistencies suggest that a more fruitful direction is to identify the minimal physical principles underlying quantum behaviour and to understand how new theories, distinct from quantum theory itself, may fit these principles. 

\textit{Acknowledgements}: This research was made possible by the generous support of the Gordon and Betty Moore Foundation, the Eutopia Foundation and the Conjecture Institute.

\end{document}